\documentclass[]{interact}

\usepackage{epstopdf}
\usepackage{subfigure}

\usepackage[numbers,sort&compress]{natbib}
\bibpunct[, ]{[}{]}{,}{n}{,}{,}

\usepackage{graphicx}
\usepackage{amssymb}
\usepackage{amsmath} 
\usepackage{color}

\newcommand{\be}{\begin{equation}}
\newcommand{\ee}{\end{equation}}
\newcommand{\bea}{\begin{eqnarray}}
\newcommand{\eea}{\end{eqnarray}}
\newcommand{\bsea}{\begin{subequations}\begin{eqnarray}}
\newcommand{\esea}{\end{eqnarray}\end{subequations}}

\newcommand{\ket}[1]{\left| #1 \right\rangle}
\newcommand{\av}[1]{\left\langle #1 \right\rangle}
\newcommand{\mb}[1]{\mathbf{#1}}
\newcommand{\rr}{\mathbf{r}}

\newcommand{\spinor}[1]{\left(
\begin{array}{c}
#1_{+} \\
#1_{0} \\
#1_{-}
\end{array}
\right)}
	
\newcommand{\spinorr}[1]{\left(
\begin{array}{c}
#1_{1} \\
#1_{0} \\
#1_{-1}
\end{array}
\right)}

\begin{document}

\articletype{ARTICLE TEMPLATE}

\title{Unified way for computing dynamics of Bose-Einstein condensates and degenerate Fermi gases}

\author{
\name{
K.~Gawryluk\textsuperscript{a}\thanks{CONTACT K.~Gawryluk. Email: k.gawryluk@uwb.edu.pl}, 
T. Karpiuk\textsuperscript{a},
M. Gajda\textsuperscript{b},
K. Rz\c{a}\.zewski\textsuperscript{c},
and M. Brewczyk\textsuperscript{a}
}
\affil{
\textsuperscript{a}Faculty of Physics, University of Bialystok, ul. Cio{\l}kowskiego 1L, 15-245 Bia{\l}ystok, Poland;\\
\textsuperscript{b}Institute of Physics PAN, Al. Lotnik{\'ow} 32/46, 02-668 Warsaw, Poland;\\
\textsuperscript{c}Center for Theoretical Physics PAN, Al. Lotnik{\'o}w 32/46, 02-668 Warsaw, Poland
}
}

\maketitle

\begin{abstract}
In this work we present a very simple and efficient numerical scheme which can be applied to study the dynamics of bosonic systems like, for instance, spinor Bose-Einstein condensates with nonlocal interactions but equally well works for Fermi gases. The method we use is a modification of well known Split Operator Method (SOM). We carefully examine this algorithm in the case of $F=1$ spinor Bose-Einstein condensate without and with dipolar interactions and for strongly interacting two-component Fermi gas. Our extension of the SOM method has many advantages: it is fast, stable, and keeps constant all the physical constraints (constants of motion) at high level.
\end{abstract}


\section{Introduction}
\label{Introduction}

The first dilute atomic Bose-Einstein condensates (BEC) were realized experimentally in 1995 by the groups of: E. Cornell and C. Wieman for rubidium \cite{Cornell},  W. Ketterle for sodium \cite{Ketterle}, and R. Hulet for lithium atoms \cite{Hulet1,Hulet2}. In these experiments, magnetic moments (associated with the electrons' and nucleus' spins) of very low energy atoms followed the external trapping magnetic field. Because of spin-spin interactions the projections of magnetic moments can change and thus atoms are no longer kept by the magnetic trap and they escape, so eventually only one component gas remains trapped (with frozen spin degree of freedom), and is described by the scalar wave function \cite{Pethick}. 

After experiments with optical traps the atoms' spin degree of freedom is not constrained to a single component only  \cite{optical_traps} what allows to study a spin dynamics due to interactions. But in this case the condensate wave function is no longer a single scalar function. It has now $2F+1$ components describing condensates of atoms of all possible spin projections of the total atom spin  $F$. Such a system is known as a spinor condensate \cite{Ueda}. 

Soon after the Bose-Einstein condensate was achieved experimentalists successfully cooled atomic Fermi gas below the degeneracy temperature \cite{Jin}. Since at very low temperatures the contact interactions in a single-component spin-polarized Fermi gas are excluded by the statistics, the experimental realization of quantum degeneracy requires trapping of two kinds of atoms. Therefore, in the case of fermions the system also becomes multicomponent. In this work we describe an efficient method for solving the dynamics of both spinor condensates and two-component Fermi gases. 

In addition to the short-range interactions between atoms we include in our analysis also the long-range dipolar interactions between magnetic moments of atoms. To study the dynamics of such a system of cold bosonic atoms we apply the mean-field approximation. For fermions, a hydrodynamical approach including the gradient corrections \cite{WeiKir1,WeiKir2} to the standard Thomas-Fermi approximation \cite{TFappr1,TFappr2}, followed by the inverse Madelung transformation \cite{Madelung} is employed. From the numerical point of view it means that we must solve the system of nonlinear partial integro-differential set of equations. This is a demanding task. Fortunately, an advantage is that the set of equations under consideration has some constants of motion. Then the algorithm which is developed can be verified against how well these quantities are preserved during the evolution. Of course, there exist algorithms belonging to the very fast developing group of structure-preserving numerical methods (called often as geometric numerical integration methods) which conserve such quantities inherently, by construction \cite{Lubich,Furihata}. However, these algorithms, to the best of our knowledge, were developed just for ordinary differential equations \cite{Lubich} or for only certain partial differential equations mainly in a one-dimensional space \cite{Furihata}. Our task is obviously much more complex. In such a case the procedure of checking the conservation of constants of motion is the only one (the other could be the comparison with existing analytical solutions which is not, however, our case) enhancing our confidence in the correctness of the algorithm. 

The paper is organized as follows. We start with bosons in Section \ref{Description}, where we introduce the Gross-Pitaevskii (GP) equation for the scalar condensate and extend it to a spinor BEC. In Section \ref{SOMscalar} we briefly describe a well known version of SOM for the case of a single GP equation. Next, we generalize this method to the spinor case (Sections \ref{SOMspinor}) and study its performance, in particular the conservation of constants of motion without dipolar (Section \ref{accuracy_tests}) and including dipolar interactions (Sections \ref{dipolar_int},\ref{numerical_results}). Then we discuss the way we treat the system of degenerate fermionic atoms (Section \ref{Fermi_gases}). We show that within the widely accepted approximations for Fermi systems their dynamics can be efficiently studied with the help of SOM method. We conclude in section \ref{conclusions}.

\section{Description of the condensate -- a mean-field approximation }
\label{Description}

We start with bosonic systems. An extremely useful approach to describe the ultra-low temperatures' properties of the Bose gas is a mean-field approximation. In this approximation we totally neglect the quantum fluctuations and substitute the field operator $\hat{\psi}(\rr)$ ($\hat{\psi}^{\dagger}(\rr)$) by a complex function $\psi(\rr)$ ($\psi^*(\rr)$) which we call the condensate wavefunction or the order parameter \cite{Pethick,Leggettreview}. This kind of treatment can be justified \cite{Dalfovo} and agrees well with many experiments \cite{mean_field}. The condensate wavefunction, $\psi(\rr)$, fulfills the well known Gross-Pitaevskii equation \cite{GP1,GP2}
\begin{equation}
\label{GPeq}
i \hbar \frac{\partial}{\partial t} \psi(\rr) = ({\cal{H}}_0+{\cal{H}}_c+{\cal{H}}_d)\, \psi(\rr)  \;.
\end{equation}
The single-particle Hamiltonian ${\cal{H}}_0$ describes the contribution from the kinetic and potential energies and equals  ${\cal{H}}_0=-\frac{\hbar^2}{2m}\nabla^2+V_{trap}$, where $m$ is a mass of an atom and $V_{trap}=\frac{1}{2}m\omega^2(x^2+y^2+z^2)$ is a trapping potential (without any loss of generality we assume here a spherically-symmetric trap). ${\cal{H}}_{c}$ is the Hamiltonian related to the contact interactions which for a scalar condensate equals ${\cal{H}}_{c}=g \psi^*(\rr) \psi(\rr)$ with the constant $g=4\pi \hbar ^2a_s/m$ characterizing the atom-atom interaction ($a_s$ is the scattering length -- the parameter which is sufficient to describe the low temperature collisions \cite{Pethick}).

A condensate of $F=1$ atoms is described by the spinor wavefunction instead of the scalar one 
\be
\psi(\rr)=(\psi_1(\rr), \psi_0(\rr), \psi_{-1}(\rr))^T,
\ee
where the component wavefunction $\psi_i(\rr)$ ($i=1,0,-1$) describes atoms in the hyperfine state $\ket{F,i}$. Now, the contact interaction part of the Hamiltonian in Eq. (\ref{GPeq}) reads 
\be
{\cal{H}}_c=
c_0\, \psi^\dagger(\mathbf{r})  
\psi(\mathbf{r}) 
+c_2 \left(\psi^\dagger(\mathbf{r})  
\mathbf{F} \psi(\mathbf{r})\right)  \cdot \mathbf{F},
\ee
where $c_0=4\pi \hbar^2(2a_2+a_0)/3m$ and $c_2=4\pi \hbar^2(a_2-a_0)/3m$ determine the strength of spin-preserving and spin-changing collisions, respectively \cite{c0c21,c0c22} (with $a_{0,2}$ being the s-wave scattering lengths for the total spin of colliding atoms equal to 0 and 2, respectively \cite{Ueda}), and $\mb{F}$ is a spin-one vector built of $F=1$ spin matrices.  We use standard definition of $\mb{F}=(F_x,F_y,F_z)$ with  
\be
F_x=\frac{1}{\sqrt{2}}
\left(
\begin{array}{ccc}
0 & 1 & 0 \\
1 & 0 & 1 \\
0 & 1 & 0
\end{array}
\right),
F_y=\frac{1}{\sqrt{2}}
\left(
\begin{array}{ccc}
0 & -i & 0 \\
i & 0 & -i \\
0 & i & 0
\end{array}
\right),
F_z=
\left(
\begin{array}{ccc}
1 & 0 & 0 \\
0 & 0 & 0 \\
0 & 0 & -1
\end{array}
\right).
\label{macierzeSpinoweF1}
\ee
The term ${\cal{H}}_c$ is responsible for the spin dynamics, using matrix notation it equals 
\be
{\cal{H}}_c=
\left(
\begin{array}{ccc}
{\cal{H}}_{c11} & {\cal{H}}_{c10} & 0 \\
{\cal{H}}_{c10}^{\star} & {\cal{H}}_{c00} & {\cal{H}}_{c0-1} \\
0 & {\cal{H}}_{c0-1}^{\star} & {\cal{H}}_{c-1-1}
\end{array}
\right),
\label{hamiltonianKontaktowy}
\ee
with definitions
\begin{eqnarray}
&&{\cal{H}}_{c11} = (c_0+c_2)\psi_1^* \psi_1+(c_0+c_2) \psi_0^* \psi_0+(c_0-c_2) \psi_{-1}^* \psi_{-1} \nonumber \\
&&{\cal{H}}_{c00} = (c_0+c_2)\psi_1^* \psi_1+c_0 \psi_0^* \psi_0+(c_0+c_2) \psi_{-1}^* \psi_{-1}     \nonumber   \\ 
&&{\cal{H}}_{c-1-1} = (c_0-c_2) \psi_1^* \psi_1+(c_0+c_2) \psi_0^* \psi_0+(c_0+c_2) \psi_{-1}^* \psi_{-1} \nonumber \\
&&{\cal{H}}_{c10}=c_2 \psi_{-1}^* \psi_0  \,, \,\,\,\,\,\,\,\,\,\,  
  {\cal{H}}_{c0-1}=c_2 \psi_{0}^* \psi_1  \,, \,\,\,\,\,\,\,\,\,\, 
  {\cal{H}}_{c1-1}=0 \,.
\end{eqnarray}    

Finally, the third term in Eq. (\ref{GPeq}) is related to the dipolar interactions and is written as
\begin{equation}
{\cal{H}}_{d} = \int d^{\,3}r'\, \psi^\dagger({\bf r'}) \, V_d ({\bf r} - {\bf r'})\, \psi({\bf r'}) \,,
\label{dipolar}
\end{equation}
where the interaction energy of two atoms with magnetic moments $\gamma \boldsymbol{F}_1$ and $\gamma \boldsymbol{F}_2$ ($\gamma$ is a gyromagnetic coefficient), positioned at ${\bf r}$ and ${\bf r'}$ is
\begin{eqnarray}
V_d ({\bf r}, {\bf r'}) = \gamma^2 \frac{\boldsymbol{F}_1 \, \boldsymbol{F}_2}{|\bf r-\bf r'|^3}-3
\gamma^2 \frac{
[\boldsymbol{F}_1 \, (\bf r-\bf r')] \,
[\boldsymbol{F}_2 \, (\bf r-\bf r')]}
{|\bf r-\bf r'|^5}   \,.
\label{ddi}
\end{eqnarray}  
As can be seen from (\ref{ddi}) the spin projection of a pair of colliding atoms can change at most by $2$, while the spin projection of a single atom changes maximally by $1$. Therefore, the matrix ${\cal{H}}_{d}$ gets tridiagonal with nonvanishing elements on diagonal equal to ${\cal{H}}_{d\zeta\zeta}=\zeta {\cal{H}}_{d11}$ and off-diagonal elements given by ${\cal{H}}_{d\zeta,\zeta-1} \! = \! \sqrt{\!(4-\zeta) (3+\zeta)/12}\, {\cal{H}}_{d10}$ ($\zeta=1,0,-1$). 

Hence, the equation of motion for the $F=1$ spinor condensate in a matrix form looks as follows
\begin{multline} 
i\hbar \frac{\partial}{\partial t} 
\spinorr{\psi}
= \\
\left(
\begin{array}{ccc}
{\cal{H}}_0 + {\cal{H}}_{c11} + {\cal{H}}_{d11}  & {\cal{H}}_{c10} + {\cal{H}}_{d10} & 0 \\
{\cal{H}}_{c10}^* + {\cal{H}}_{d10}^* & {\cal{H}}_0 + {\cal{H}}_{c00} & {\cal{H}}_{c0-1} + {\cal{H}}_{d10} \\
0 & {\cal{H}}_{c0-1}^* + {\cal{H}}_{d10}^* & {\cal{H}}_0 + {\cal{H}}_{c-1-1} - {\cal{H}}_{d11}
\end{array}
\right)
\spinorr{\psi}.
\label{spinoroweRownaniaRuchu} 
\end{multline}
In the simplest, i.e. scalar, case it becomes the Gross-Pitaevskii equation supplemented by the nonlocal term due to the dipolar interactions \cite{Goral1,Goral4}
\bea 
i\hbar \frac{\partial}{\partial t} \psi(\rr) & = &
\left(-\frac{\hbar^2}{2m}\nabla^2+V_{trap}(\rr) + g|\psi(\rr)|^2 \right. \nonumber \\
& + &
\left. \gamma^2 \int d^3r' \left[ 
\frac{1}{|\mathbf{r}-\mathbf{r}'|^3}-3\frac{(z-z')^2}{|\mathbf{r}-\mathbf{r}'|^5} 
\right] |\psi(\mathbf{r}')|^2 \right) \psi(\rr) . 
\label{scalar_GPeq}
\eea
It is worth to notice that the above equation has the form of the Schr\"odinger equation with additional nonlinear and nonlocal terms which take into account the mean-field and dipolar energies of interacting bosons.

One component Gross-Pitaevskii equation without and with additional dipolar interactions, Eq. (\ref{scalar_GPeq}), has been already attempted by several authors \cite{GPdipolar1,GPdipolar2,GPdipolar3,GPdipolar4,WangDipSkal,Muruganandam,Vudragovic,KishorKumar}. Also, the spinor version of the GP equation, Eq. (\ref{spinoroweRownaniaRuchu}), but without including dipole-dipole term ${\cal{H}}_d$ has been studied \cite{GPdipolar_no1,GPdipolar_no2,WangF1}. Here, for the first time we investigate the quality of a numerical algorithm while atomic transfer between different spin components, due to dipolar interactions, is allowed. Introducing the dipolar interactions into the spinor condensate is by no means a trivial extention. In fact, the resulting atomic flow between components corresponds to the famous Einstein-de Haas effect \cite{EinsteinHass}. From the numerical point of view it means that another constant of motion emerges in this case, which is the projection of the total angular momentum. Any proposed algorithm should be checked for the quality of the conservation of this new constant of motion. 

In the rest of this paper we will be often using the oscillatory units, in which the units of time, length, energy, coupling constant, and gyromagnetic coefficient are given by $\tau=1/\omega$, $l=\sqrt{\frac{\hbar}{m \omega}}$, $E=\hbar \omega$, $E l^3$, and $\sqrt{\omega l^3}$, respectively.

\section{Split Operator method -- a scalar version}
\label{SOMscalar}

In this section, just for completeness, we briefly present SOM for a scalar order parameter (for more details see Refs. \cite{SOMsolution,SOMcomparison,SOMext,Bao13}).  
Our goal is to  find  $\psi(t)$ satisfying (\ref{scalar_GPeq})
with the initial condition at $t=0$ given by $\psi(0)$. To this end we divide the time $t$ into $N$ intervals $\Delta t$ such that $t=N \Delta t$. Then, assuming that $\Delta t$ is small, we can write (with the first order of accuracy in $\Delta t$)
\be
\psi(t)=\prod_{n=0}^{N-1}\exp\big( -i \Delta t H(n\Delta t) \big) \psi(0),
\ee
where $H(t)$ is the right hand-side operator from (\ref{scalar_GPeq}). In particular one has
\be
\psi(t+\Delta t)=\exp \left[ -i
\left(-\frac{1}{2} \nabla^2+\frac{1}{2} r^2+g|\psi(t)|^2 + V_{dip}(t) \right)
\Delta t \right] \psi(t),
\ee
 where for convenience only the dipolar contribution was denoted by $V_{dip}$.
Using  $e^{A+B} \approx e^A e^B$  from the Baker-Hausdorff theorem\footnote{
The \textit{Baker-Hausdorff} theorem reads
$
e^{A+B} = e^A e^Be^{-\frac{1}{2}[A,B]} \mbox{   if }[A,[A,B]]=0 \mbox{ and } [B,[A,B]]=0.
\label{twHausdorfBakera}
$
In our case the definitions of a $A$ and $B$ are $A=i\frac{\Delta t}{2}  \nabla^2,\;B=-i \Delta t\left(\frac{1}{2} r^2+g|\psi(t)|^2+V_{dip}(t)\right)$
and are $\Delta t$-dependent, and $e^{-\frac{1}{2}[A,B]} \propto I$ (because it's $\propto \Delta t^2$), so one can write
$e^{A+B} \approx e^A e^B.$ } we obtain
\be
\psi(t+\Delta t) \approx \exp \left[ i\frac{\Delta t}{2}  \nabla^2 \right]
\exp \left[ -i \Delta t\left(\frac{1}{2} r^2+g|\psi(t)|^2 + V_{dip}(t) \right) \right] \psi(t).
\label{ewolucja_o_dt}
\ee
The above approximation is getting justified when $\Delta t\ll 1$ (because commutator of the $A$ and $B$ operators appearing in the Baker-Hausdorff theorem is $\Delta t$-dependent, giving $(\Delta t)^2$ dependency in total). The right-hand side of Eq. (\ref{ewolucja_o_dt}) is just the Lie-Trotter splitting \cite{Lubich1} applied to the GP equation (\ref{scalar_GPeq}). It is easy to improve the convergence of the above scheme by symmetrizing it with respect to one of the operators: $A$ or $B$. It can be shown that the one-step formula of the form
\be
\psi(t+\Delta t) \approx \exp \left[ i\frac{\Delta t}{4}  \nabla^2 \right]
\exp \left[ -i \Delta t\left(\frac{1}{2} r^2+g|\psi(t)|^2 + V_{dip}(t) \right) \right] 
\exp \left[ i\frac{\Delta t}{4}  \nabla^2 \right]   \psi(t)
\label{ewolucja_Strang}
\ee
is of the second order in time. It is called the Strang splitting \cite{Lubich1}. The operator $\exp (i (\Delta t /4) \nabla^2 )$ does not depend on time. Therefore, when two successive time steps of the evolution scheme (\ref{ewolucja_Strang}) are accomplished the neighbouring operators $\exp (i (\Delta t /4) \nabla^2 )$  merge and take the form of $\exp (i (\Delta t /2) \nabla^2 )$ .

 Hence, effectively a single time-step $\Delta t$ evolution (\ref{ewolucja_Strang}) can be split into two steps (only the first and the last steps in the sequence (\ref{ewolucja_Strang}) should be treated differently). First, we define an auxiliary function
\be
\psi_1(t)=\exp \left[ -i \Delta t\left(\frac{1}{2} r^2+g|\psi(t)|^2+V_{dip}(t) \right) \right]\psi(t),
\ee
which we transform to the k-space using the Fourier transform
\be
\tilde{\psi}_1=\mathcal{F}[\psi_1],
\label{somEk_1}
\ee
and then we evaluate  the  expression: $\psi_2=\exp [i \frac{\Delta t}{2} \nabla^2] \psi_1$. In \mbox{$k$-space} it becomes a simple multiplication:
\be
\tilde{\psi}_2=\exp \left[ -\frac{i \Delta t}{2}k^2\right] \tilde{\psi}_1.
\label{somEk_2}
\ee
Doing this, we compute the order parameter at time $t+\Delta t$, but it is still in momentum space. So, the last step is to return to the coordinate space (with the help of the inverse Fourier transform)
\be
\psi(t+\Delta t)=\mathcal{F}^{-1}[\tilde{\psi}_2].
\label{somEk_3}
\ee

\section{Split Operator method -- a spinor version}
\label{SOMspinor}

In this section we introduce our extension to the SOM. For $F=1$ atoms the order parameter $\psi$ consists of three components $\psi=(\psi_{+},\psi_0,\psi_{-})$ which satisfies the equation similar to the Eq. (\ref{spinoroweRownaniaRuchu})
\be
i \frac{\partial}{\partial t}\spinor{\psi}=\big(-\frac{1}{2}\nabla^2 +V(\mb{r},t)\big)\spinor{\psi}.
\label{spinorGP}
\ee
Here by $V(\mb{r},t)$  we denoted the sum of the trapping potential and the nonlinear terms appearing in the right-hand side of Eq. (\ref{spinoroweRownaniaRuchu}) describing the contact interactions between atoms. Then we use the scheme described in the previous section  and obtain
\be
\spinor{\psi}(t+\Delta t) \approx \exp 
\left[ i\frac{\Delta t}{2}  \nabla^2 \right]
\exp \Big[ -i \Delta t V(\mb{r},t) \Big]
\spinor{\psi}(t).
\ee
Again, we introduce an auxiliary function $\psi_1$ (this time as a $3$-component vector), which we are going to transform to the momentum space. But first we need a careful treatment of a `potential term' $\exp{[-i \Delta t V(\mb{r},t)]}$ since $V(\mb{r},t)$ is a matrix. The evolution in the position space requires calculation of the following expression:  
\be
\exp[-i \Delta t V(\mb{r},t)]\spinor{\psi}(t).
\label{spinorowyV}
\ee
To calculate (\ref{spinorowyV}) 
we bring the $V(\mb{r},t)$ matrix to the diagonal form:
\be
\label{PVP-1}
V=P D P^{-1},\;\;\;
D=\left(
\begin{array}{ccc}
\lambda_{+} & 0 & 0 \\
0 & \lambda_0 & 0 \\
0 & 0 & \lambda_{-}
\end{array}
\right),
\ee
which, after utilizing the expression $\exp x= \sum_n\frac{x^n}{n!}$, gives 
\be
\exp[-i \Delta t V] \! \spinor{\psi}\!(t)=\left(\! 1-i \Delta t P D P^{-1} + \frac{1}{2!}(-i \Delta t P D P^{-1})^2+\ldots \! \right)\!\spinor{\psi}\!(t).
\ee
The above expansion greatly simplifies after applying $V^n=P D^n P^{-1}$ formula. One gets
\be
\exp[-i \Delta t V] \spinor{\psi}(t)=P \left( 1-i \Delta t D + \frac{1}{2!}(-i \Delta t D)^2+\ldots \right)P^{-1}\spinor{\psi}(t).
\ee
Next we can collect terms with $D^n$ back to the exponential form
\be
\exp[-i \Delta t V] \spinor{\psi}(t)=P
\exp[-i \Delta t D]P^{-1}
\spinor{\psi}(t).
\ee
One can see now that we converted a problem of calculating (\ref{spinorowyV}) to calculation of the exponential function of a $D$ matrix which is diagonal. So, we have 
\be
\label{wynik_expV}
\exp[-i \Delta t V]\spinor{\psi}(t)=P
\left(
\begin{array}{ccc}
e^{-i \lambda_{+} \Delta t} & 0 & \\
0 &e^{-i \lambda_{0} \Delta t} & 0 \\
0 & 0 & e^{-i \lambda_{-} \Delta t}
\end{array}
\right) P^{-1}\spinor{\psi}(t).
\ee
The idea of calculating (\ref{spinorowyV}) by diagonalizing $V(\mb{r},t)$ is the essence of our extension of the original SOM\footnote{In fact we copied the original idea of the SOM: to calculate the evolution due to the kinetic part of the Hamiltonian one has to switch to the basis, in which the Laplace operator is diagonal, i.e., which is the momentum space. Because in the momentum space the differentiation becomes a simple multiplication by $-i\mathbf{k}$, the evolution due to the kinetic term (which includes the second derivative in position space) becomes very easy to compute - instead of the second derivative one has to multiply Fourier components by $\exp[-k^2\Delta t/2]$. In the spinor version we have to do a similar trick in the position space: to facilitate calculations of the evolution according to the potential energy, first we go to the basis in which the position dependent part of the Hamiltonian $V(\mb{r},t)$ is diagonal, and only then we use the plane wave basis for evolution due to a kinetic part.}.       A reader can think about (\ref{wynik_expV}) as a result of taking infinitely many terms in the Taylor expansion of the evolution operator
\be
\label{taylor_expV}
\exp[-i \Delta t V] \! \spinor{\psi} \! (t)=
\left(\!
1-i\Delta t V+\frac{(-i \Delta t V)^2}{2!}+\frac{(-i \Delta t V)^3}{3!}+\ldots \!
\right) \! \spinor{\psi}\!(t).
\ee
Utilizing (\ref{wynik_expV}) allows to overcome many problems: 1) we do not need to worry about the number of terms one should use to calculate right-hand side of (\ref{taylor_expV}) to achieve a good enough accuracy, and 2) diagonalization of small matrices ($3\times 3$ for $F=1$ atoms or $7\times 7$ for $F=3$ atoms, like $^{52}$Cr atoms) is significantly more efficient than calculating the right-hand side of (\ref{taylor_expV}), even assuming that only a few terms are taken into account.

Once we successfully calculated the evolution due to the potential part of the Hamiltonian, we can follow remaining SOM steps as described in the previous section: first we use (\ref{wynik_expV}) to calculate  an auxiliary spinor function 
\be
\psi_1 = \exp[-i \Delta t V]\spinor{\psi}(t).
\ee
Next we move to the $k$-space with $\psi_1$
\be
\tilde{\psi}_1=\mathcal{F}[\psi_1],
\ee
in order to calculate the evolution according to the kinetic energy part in the Hamiltonian
\be
\tilde{\psi}_2=\exp \left[ -\frac{i \Delta t}{2}k^2\right] \tilde{\psi}_1.
\ee
Finally we come back to the coordinate space
\be
\psi(t+\Delta t)=\mathcal{F}^{-1}[\tilde{\psi}_2].
\ee
This sequence of steps propagates the initial wave function over the time interval $\Delta t$:  $\psi(t) \rightarrow \psi(t+\Delta t)$.

It is clear from the above scheme, that SOM in a spinor version for $F=1$ atoms requires diagonalization of a $3 \times 3$ matrix at every spatial point on the grid and $3 \times 2 = 6$ Fourier transforms -- a pair of forward and backward ones for each component of $\psi_i$ ($i$ goes from $1$ to $3$).

\section{Accuracy tests for spinor condensates}
\label{accuracy_tests}

We will present accuracy tests for real time propagation only, having already calculated the initial state by the imaginary-time propagation technique \cite{imag_time} applied to our extension to the SOM. We focus on real time propagation because the GP equation (\ref{spinoroweRownaniaRuchu}) conserves the energy
\be
\label{energy_constr}
\av{E}=const,
\ee
the total norm
\be
\label{tot_norm_constr}
N_{tot}=N_{+}+N_0+N_{-}=const
\ee
with $N_i=\int |\psi_i(\textbf{r})|^2d^3r\; (i={+,0,-})$, and, assuming the dipolar interactions are neglected, the magnetization
\be
\label{magnetiz_constr}
N_{+}-N_{-}=const\,.
\ee
The expressions (\ref{energy_constr}), (\ref{tot_norm_constr}) and (\ref{magnetiz_constr}) are not any constraints for imaginary time-evolution. The energy decreases monotonically in every step of imaginary-time evolution till the ground state is reached  \cite{imag_time}. The norm decreases as well and therefore after each single step of computations the wave function is normalized. Any algorithm attempting to solve the GP equation (\ref{spinoroweRownaniaRuchu}) should be verified with respect to how these constants of motion are preserved during the evolution.

To test the conservation of the constants of motion (\ref{energy_constr}-\ref{magnetiz_constr}) (the case including the dipolar interactions will be discussed in the next section) we chose a system of $3\times 10^4$ $^{87}$Rb atoms, with $a_0=5.387$ nm and $a_2=5.313$ nm (according to Ref. \cite{a0a2}) as contact interaction parameters, confined in a pancake trap ($\omega_x = \omega_y=2\pi \times 100$ Hz, $\omega_z=2\pi \times 2000$ Hz). We start with all the atoms in $m_F=0$ component and monitor the evolution of the system. Since the contact interactions (\ref{hamiltonianKontaktowy}) allow for the transfer of atoms from the initial state to the $m_F=\pm 1$ states, one could expect the appearance of spin dynamics. This dynamics was already investigated in \cite{termalizacjaF1} but here we will focus mainly on the numerical aspects, refereeing the reader to Ref. \cite{termalizacjaF1} for a deeper understanding of the physical background. 

\begin{figure}[ht]
\centering
\includegraphics[scale=0.43]{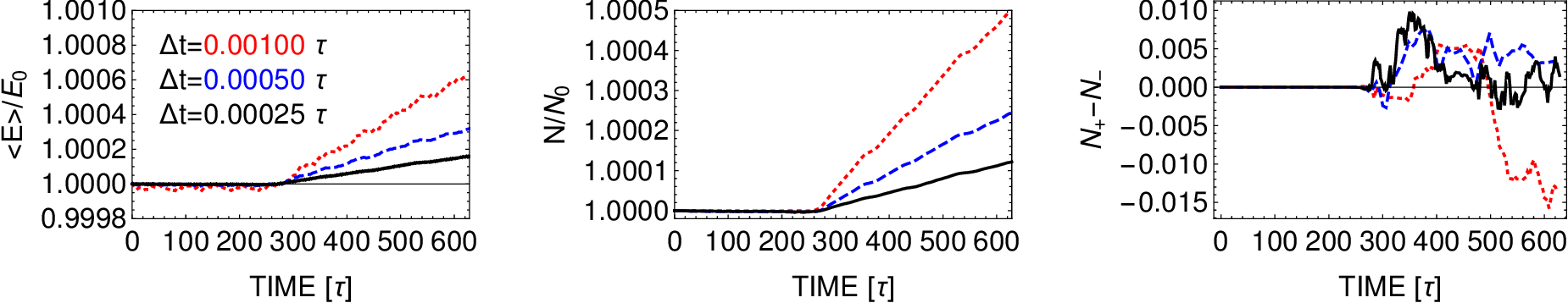}
\caption{Illustration of the conservation of the total energy, the total norm, and the magnetization of a spin-1 atoms system. $E_0$ and $N_0$ are the energy and the number of atoms at $t=0$, respectively. Details of the system are given in the text. Colors depict different time steps used in simulations: $\Delta t=0.00025 \tau$ (black), $\Delta t=0.0005 \tau$ (blue), and $\Delta t=0.001\tau$ (red). Time unit equals $\tau=1.59155$ ms.}
\label{fig_constants_dt}
\end{figure}

We performed fully 3D simulations on a Cartesian grid of $2^52^52^4$ points, with the spatial steps equal to $\Delta x=\Delta y= 0.6, \Delta z= 0.1$ in oscillatory units (we have chosen $\hbar \omega_x$ as a unit of energy, thus the oscillatory unit of length equals $l=1.07961 \mu$m). Fig. \ref{fig_constants_dt} shows the conservation of the total energy, the total number of atoms, and the magnetization of a spin-1 gas for different time steps $\Delta t$. It is not surprising that the biggest time step leads to the worst conservation of the energy and the norm. This behavior is expected since $\Delta t$ plays the crucial role in the SOM (compare $\Delta t$ dependence of commutators in the Baker-Hausdorff theorem (\ref{twHausdorfBakera})). Even in the worst case presented here the energy is preserved with accuracy better than 0.1\%, which is acceptable in many simulations. It is worth to notice that going into smaller $\Delta t$ (for example decreasing it by a factor of $2$) significantly improves accuracy (by a factor of 2). It is important that we can easily control this accuracy by modifying  $\Delta t$. Fig. \ref{fig_constants_dt} also shows that the spinor SOM conserves the magnetization (\ref{magnetiz_constr}) which is expressed here in absolute numbers. Let us remind that the total number of atoms in the  system is $N_{tot}=30000$. 

Fig. \ref{fig_constants_dt} shows a few interesting features: 1) conservation of the total energy starts to be worse at particular time ($t \approx 280 \tau$), and 2) we note an almost linear behavior of $\av{E}/E_0$ curves -- so one might be worried about conservation of the energy for longer evolution time ($t>600 \tau$). To clarify the second issue we would like to remind the reader that we are dealing with ultracold atoms and that the typical lifetime of the condensate is of the order of a few seconds \cite{Pethick,Dalfovo,Leggettreview}. That is why there is no need to continue the evolution for more than here ($1$ s). Let us note however, that extending the total time by a factor of $5$  (up to $5$ seconds, which is a large evolution time from the point of view of real experiments) we get the total energy conserved  with  $1\%$ accuracy (for $\Delta t=0.0005 \tau$) or even better -- with $0.5\%$ accuracy (in case of $\Delta t=0.00025 \tau$). That gives us an excellent level of conservation of the total energy, and moreover - we can control it by a proper choice of $\Delta t$. In Tab. \ref{spinortab} we compare the cumulative errors for different spatial grids. The case (B) corresponds to the results already presented in Fig. \ref{fig_constants_dt}, whereas the case (A) stands for twice larger grid but with twice smaller spatial steps, i.e. both grids cover the same space volume. Tab. \ref{spinortab} clearly shows better conservation of discussed constants of motion for finer grids and smaller time steps.

\begin{table}[!ht]
\centering
\begin{tabular}{c|c|c|c}\hline
$\Delta$t [$\tau$] & $\frac{1}{T}\int\! dt |\frac{\av{E}-E_0}{E_0}|\; [\times 10^{-5}]$ & $\frac{1}{T}\int\! dt
|\frac{N-N_0}{N_0}|\; [\times 10^{-5}]$  & $\int\! dt\; |N_{+}-N_{-}|$  [$\tau$] \\
   & (A)\;\;\;\;\;\;\;(B) & (A)\;\;\;\;\;\;\;(B)  &
(A)\;\;\;\;\;\;\;\;(B)  \\
\hline
0.001\phantom{00}  & $15.2587\;\;70.2591$     & $11.5962\;\;13.6332$  & 
$0.96882\;\;2.02645$ \\
0.0005\phantom{0}  & $7.49158\;\;31.0140$     & $5.26571\;\;5.60201$  & 
$0.70001\;\;1.37984$ \\
0.00025            & $4.00000\;\;17.1502$     & $2.89349\;\;3.00013$  & 
$0.62613\;\;0.95889$\\
\hline
\end{tabular}
\caption{Cumulative errors: (A) -- for the grid with $2^62^62^5$ points and the spatial steps twice smaller than those used to produce results presented in Fig. \ref{fig_constants_dt}; (B) -- for the grid with $2^52^52^4$ points and the spatial steps as in Fig. \ref{fig_constants_dt}. In both cases $T=628\tau$.}
\label{spinortab}
\end{table}

\begin{figure}[ht]
\centering
\includegraphics[scale=0.43]{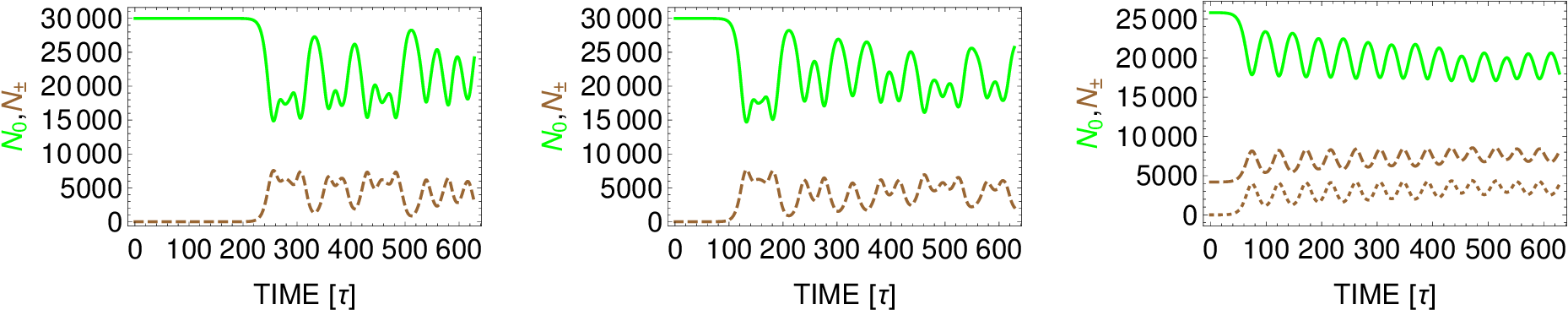}
\caption{Time evolution of the number of atoms in each component for different values of initial seed. Left frame: $N_{+}(0)=N_{-}(0)=10^{-12}$. Middle frame: $N_{+}(0)=N_{-}(0)=10^{-1}$. Right frame: $N_{+}(0)=10^{-12}$, $N_{-}(0)=4197$. Other parameters: $\Delta t=0.0005 \tau$, $2^5 2^5 2^4$ points grid. Although a large transfer of atoms to $m_F=\pm 1$ states starts at different times, the accuracy measured by a degree of conservation of the total energy and the total norm remains the same. For a deeper understanding of the physics behind we suggest to read Ref. \cite{termalizacjaF1}. }
\label{fig_seeds}
\end{figure}

What happens around $t \approx 280 \tau$ is shown in Fig. \ref{fig_seeds}, left frame. Evidently, some nontrivial dynamics is triggered -- atoms start to flow from $m_F=0$ to $m_F=\pm 1$ components. Obviously, some spin dynamics might be expected in the system since our initial state is not the lowest energy solution of the spinor configuration. Spin changing collisions between atoms in the $m_F=0$ state start to produce atoms in $m_F=\pm 1$ states. This reach spin dynamics evolves the system into the state of thermal equilibrium. This state is approached in a steplike process, see Ref. \cite{termalizacjaF1}.

Around time $t\approx 280 \tau$ a large number of atoms is transferred from initially populated $m_F=0$ component to the other states (Fig. \ref{fig_seeds}, left frame). One might wonder if this characteristic time at which the spin dynamics is triggered  depends on the initial condition. As we have already mentioned, initially we assume  $N_{tot}=30000$ atoms in $m_F=0$ component and almost zero atoms in $m_F=\pm 1$ states. To trigger the spin dynamics we need some seed in initially empty components, and, in fact, we used a seeding field in $m_F=\pm 1$ states. The seed was implemented  by choosing a complex random number at each point of the spatial grid. The seed plays a role of quantum fluctuations which are present in a real system. The quantum fluctuations are missing in the mean-field description (we have neglected all the quantum fluctuations). To see the spin dynamics   \footnote{By spin dynamics we understand the situation with non-negligible transfer of atoms from one component to the other. From Eq. (\ref{spinoroweRownaniaRuchu}) it follows that $\psi_1$ satisfies the equation
$i \frac{\partial \psi_1}{\partial t}=({\cal{H}}_0+{\cal{H}}_{c11})\psi_1 + {\cal{H}}_{c10}\psi_0=({\cal{H}}_0+(c_0+c_2)(|\psi_1|^2|\psi_0|^2)+(c_0-c_2)|\psi_{-1}|^2)\psi_1+c_2\psi_{-1}^\star \psi_0 \psi_0$
and if  $\psi_1(0)=\psi_{-1}(0)=0$ the right hand side vanishes, i.e. the $\psi_1$ field is constant what corresponds to  a `spin frozen' situation. To account for quantum fluctuations we need some small initial seeds in the fields $\psi_1$ and $\psi_{-1}$.}     the seed must be present at $t=0$. We have checked, that changing the amount of seed (i.e. values of $N_{+}(0)$ and $N_{-}(0)$ being the total number of atoms in $m_F=+1,-1$ states, respectively, at $t=0$) results in shifting the time at which the significant transfer of atoms starts. For example for initial values $N_{+}(0)=N_{-}(0)=10^{-12}$ the spin dynamics begins around $t\approx 280 \tau$ (Fig. \ref{fig_seeds}, left frame), but for $N_{+}(0)=N_{-}(0)=10^{-1}$ it happens at $t\approx 100 \tau$ (Fig. \ref{fig_seeds}, middle frame). We have also checked the imbalanced initial conditions by taking the initial populations as $N_{+}(0)=10^{-9}$ and $N_{-}(0)=4197$. Fig. \ref{fig_seeds}, (right frame) shows how the populations depend on time (brown solid line -- $N_{+}(t)$, brown dotted line -- $N_{-}(t)$). The time at which the spin dynamics is triggered depends monotonically on the amount of the seed. What is important is that the value of the seed does not change qualitatively the dynamics - it changes only the time at which the non-trivial dynamics begins. Although the transfer of atoms starts at different times, the conservation of the total energy and the total norm (as well as the magnetization) is of the same order in all situations.

\section{Dipolar interactions}
\label{dipolar_int}

It is well known that the dipolar interactions couple the spin and the orbital motion of colliding atoms. The projection of total spin of interacting atoms can change at most by $2$ as implied by the expression (\ref{ddi}). When it happens it means that the spin goes to the orbital angular momentum of atoms. In other words, atoms changing their spin must acquire orbital motion. This is the famous Einstein-de Haas effect \cite{EinsteinHass} which has been already discussed also for the systems of ultracold atoms \cite{EdH1,Ueda1,EdH2}. It can be rigorously shown that the sum of the projections of the total spin and the total orbital angular momentum is preserved during collision. Indeed, the commutator $[V_d, L_{1z}+L_{2z}+F_{1z}+F_{2z}]$ equals zero, where $L_{1z}$ and $L_{2z}$ denote the projections of the orbital angular momenta of colliding atoms \cite{EdH1}. Therefore, assuming the external potential has an axial symmetry along the $z$ axis, the quantity $L_z+F_z$ should be conserved during the evolution according to the GP equation (\ref{spinoroweRownaniaRuchu})
\bea 
L_z+F_z = const  \,.
\eea
Any algorithm attempting to solve the GP equation (\ref{spinoroweRownaniaRuchu}) with dipolar interactions should be verified against the quality of this constant of motion.

But first we have to calculate the elements of $H_{d}$ matrix (see the formula (\ref{dipolar}) and the discussion after it). One needs, actually, to obtain only two matrix elements, ${\cal{H}}_{d 11}$
\bea 
&&{\cal{H}}_{d 11}\;(\mathbf{r}) = \gamma^2 \int d^3r' \left[ 
\frac{1}{|\mathbf{r}-\mathbf{r}'|^3}-3\frac{(z-z')^2}{|\mathbf{r}-\mathbf{r}'|^5} 
\right] \times (|\psi_1(\mathbf{r}')|^2-|\psi_{-1}(\mathbf{r}')|^2) \nonumber \\ 
&& -3 \frac{\gamma^2}{\sqrt{2}} 
\int d^3r' \frac{z-z'}{|\mathbf{r}-\mathbf{r}'|^5}[(x-x')-i(y-y')] \times 
(\psi_1^{\star}(\mathbf{r}')\psi_0(\mathbf{r}')+\psi_0^{\star}(\mathbf{r}')\psi_{-1}(\mathbf{r}'))    \nonumber \\ 
&& -3 \frac{\gamma^2}{\sqrt{2}} 
\int d^3r' \frac{z-z'}{|\mathbf{r}-\mathbf{r}'|^5}[(x-x')+i(y-y')] \times 
(\psi_0^{\star}(\mathbf{r}')\psi_1(\mathbf{r}')+\psi_{-1}^{\star}(\mathbf{r}')\psi_0(\mathbf{r}')) \nonumber \\
\label{Hd_11} 
\eea
and  ${\cal{H}}_{d 10}$
\bea
&&{\cal{H}}_{d 10}\;(\mathbf{r}) = -3 \frac{\gamma^2}{\sqrt{2}} \int d^3r'  
\frac{[(x-x')-i(y-y')](z-z')}{|\mathbf{r}-\mathbf{r}'|^5} \times 
(|\psi_1(\mathbf{r}')|^2-|\psi_{-1}(\mathbf{r}')|^2)    \nonumber \\ 
&& -\frac{3}{2} \gamma^2 \int d^3r' 
\frac{[(x-x')-i(y-y')]^2}{|\mathbf{r}-\mathbf{r}'|^5}  \times
(\psi_1^{\star}(\mathbf{r}')\psi_0(\mathbf{r}')+\psi_0^{\star}(\mathbf{r}')\psi_{-1}(\mathbf{r}')) 
\nonumber  \\ 
&& + \gamma^2 \int d^3r' \left[ 
\frac{1}{|\mathbf{r}-\mathbf{r}'|^3}-\frac{3}{2}\frac{(x-x')^2+(y-y')^2} 
{|\mathbf{r}-\mathbf{r}'|^5} \right] \times 
(\psi_0^{\star}(\mathbf{r}')\psi_1(\mathbf{r}')+\psi_{-1}^{\star}(\mathbf{r}')\psi_0(\mathbf{r}'))   \,\,.  \nonumber \\
\label{Hd_10} 
\eea
All other nonvanishing elements can be expressed in terms of ${\cal{H}}_{d 11}$ and ${\cal{H}}_{d 10}$.

Integrals appearing in the expressions for the matrix elements (\ref{Hd_11}) and (\ref{Hd_10}) are the convolutions, therefore to calculate them it is convenient to use the Fourier transform technique. This is because the Fourier transform of the convolution is the product of the Fourier transforms of the functions which are convolved. The convolutions which contribute to (\ref{Hd_11}) and (\ref{Hd_10}) convolve functions which originate from the dipolar interactions with the ones composed based on the spinor components. Since the spinor wave function evolves according to the GP equation (\ref{spinoroweRownaniaRuchu}), the Fourier transform of the part of (\ref{Hd_11}) and (\ref{Hd_10}) dependent on the spinor components is calculated numerically. On the other hand, the Fourier transform of the part which originates from the dipolar interaction is calculated analytically. The rest of this Section explains how it is done.

\begin{figure}[htb]
\centering
\includegraphics[width=7cm]{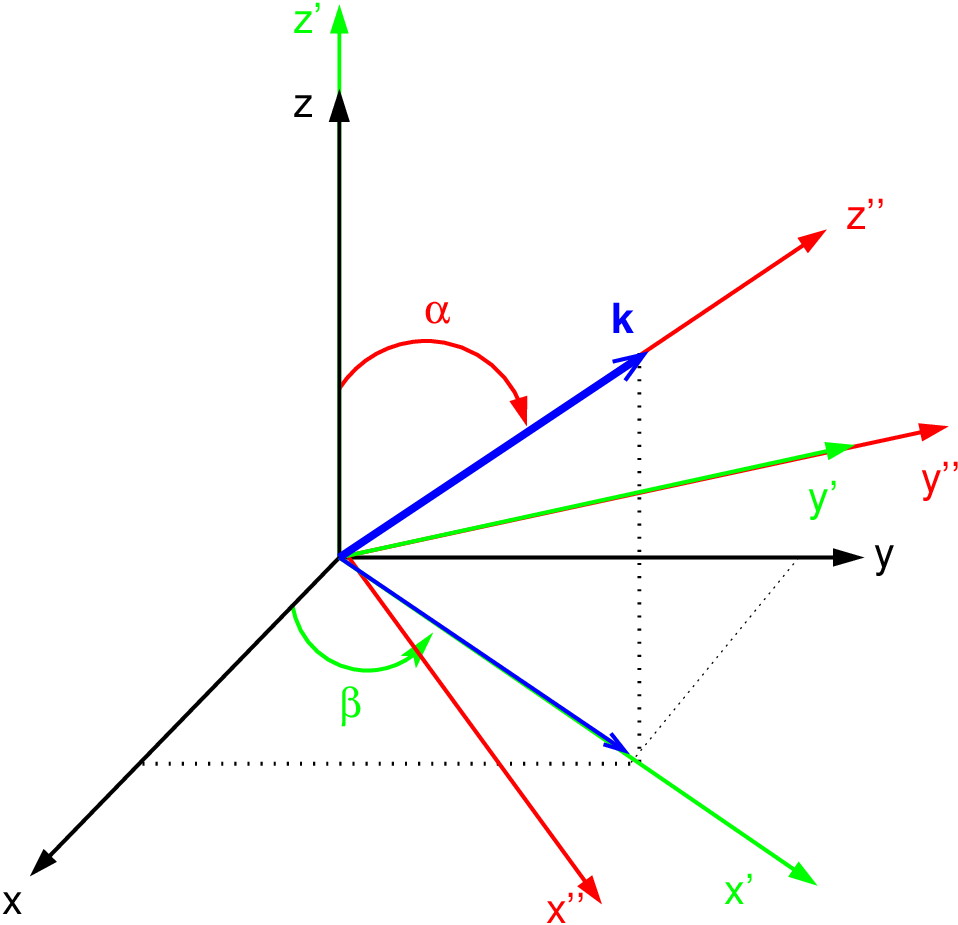}
\caption{The scheme of rotations used to calculate analytically Fourier transforms of functions which originate from dipolar interactions.}
\label{uklad}
\end{figure}

We use the following convention of the Fourier transform \cite{FFTWmanual}
\be
{\cal{F}} [ f(\mb{r}) ] = \tilde{f}(\mb{p}) =\int d^3 r\; \exp \left( +\frac{i}{\hbar} \mb{p}\cdot \mb{r} \right) f(\mb{r}).
\label{transformata_konwencja}
\ee
To evaluate (\ref{transformata_konwencja}) it is convenient to transform coordinates in such a way that vectors $\mb{p}$ and $\hat{\mb{z}}$ become parallel. This, of course, simplifies a scalar product $\mb{p}\cdot \mb{r}$ appearing in the argument of the exponential function. The required coordinate transformation is a product of two rotations: the first one is the rotation by an angle $\beta$ around $z$ axis and the second one by an angle $\alpha$ around already rotated $y$ axis (see Fig. \ref{uklad}). The final rotation matrix is given by   
\be
\left(
\begin{array}{c}
x \\
y \\
z
\end{array}
\right)=
\left(
\begin{array}{ccc}
\cos \alpha \cos \beta   & -\sin \beta     &  \sin \alpha \cos \beta \\
\cos \alpha  \sin \beta  & \cos \beta      & \sin \alpha \sin \beta  \\
-\sin \alpha                    & 0                 &  \cos \alpha
\end{array}
\right)
\left(
\begin{array}{c}
x'' \\
y'' \\
z''
\end{array}
\right).
\label{obroty_alpha_beta}
\ee
Using this transformation we set the $\mb{k}$ vector (an argument of the Fourier transform, $\mb{k}=\mb{p}/\hbar$) parallel to the $z$ axis. Definitions of used angles, from Fig. \ref{uklad}, are: $\cos \alpha = k_z/k$, $\sin \alpha = \sqrt{k_x^2+k_y^2}/k$, $\cos \beta= k_x/\sqrt{k^2-k_z^2}$, and $\sin \beta=k_y/\sqrt{k^2-k_z^2}$.

As an example we will show in detail the calculation of the ${\cal{F}} \left[ \frac{1}{|\mb{r}|^3}-3\frac{z^2}{|\mb{r}|^5} \right]$. This particular Fourier transform gets important when the single component dipolar condensate, the system investigated already in the early days of BEC \cite{Goral1,Goral2,Goral3,Goral4}, is considered. According to the definition (\ref{transformata_konwencja}) we do integrate
\be
I=\int d^3 r\; e^{i \mb{k r}}\left( \frac{1}{r^3}-3\frac{z^2}{r^5}\right)
\label{example}\centering
\ee
with the help of the transformation of coordinates given by (\ref{obroty_alpha_beta}). 
Then, going to spherical coordinates and integrating over the azimuthal angle one obtains
\be
I=\int_0^{\infty}d r\; \int_{-1}^1 dt\; e^{i k r t}
\frac{\pi}{r}\left( 1-3 \cos^2 \alpha \right) (3 t^2 -1)  \,,
\label{becareful}
\ee
where $t=\cos \theta$ and $\theta$ is the polar angle. Now, doing double integration, first over $t$ variable and then over the distance $r$ we finally arrive at
\be
I=\frac{2\pi}{3}\big(1+3 \cos(2 \alpha) \big)  \,.
\label{final}
\ee

One has to be careful in calculating the expression (\ref{becareful}). It is easy to check that the result diverges if we first integrate over $r$ variable. This is because, in fact, the integral (\ref{example}) does not converge and the regularization is required. It can be done based on the physical arguments saying that the size of the dipole is finite. Therefore, going back to (\ref{becareful}) one can first integrate it over $r$ variable within the interval $(R,\infty)$ getting
\be
I=\int_{-1}^1 dt\; \pi\, ( 1-3 \cos^2 \alpha)\, (3 t^2 -1)\, \Gamma(0,-i\,k\,R\,t) \,,
\ee
where $\Gamma$ is the incomplete gamma function. Integrating over $t$ variable and taking the limit $R\to 0$ lead us again to the result (\ref{final}).

The other Fourier transforms we treat in a similar way and obtain \cite{Swislocki} 
\bea
&&{\cal{F}} \left[ 
\frac{1}{|\mathbf{r}|^3}-3\frac{z^2}{|\mathbf{r}|^5} \right]  
= -\frac{4\pi}{3}(1-3 \cos^2\alpha)   \nonumber  \\ 
&&{\cal{F}} \left[ 
\frac{1}{|\mathbf{r}|^3}-\frac{3}{2}\frac{x^2+y^2}{|\mathbf{r}|^5} \right] 
= \frac{2\pi}{3}(1-3 \cos^2\alpha)   \nonumber  \\ 
&&{\cal{F}} \left[ 
\frac{(x-iy) z}{|\mathbf{r}|^5} \right] = -
\frac{2\pi}{3}\, e^{-i \beta} \sin 2 \alpha   \nonumber  \\ 
&&{\cal{F}} \left[ 
\frac{(x+iy) z}{|\mathbf{r}|^5} \right] = -
\frac{2\pi}{3}\, e^{i \beta} \sin 2 \alpha   \nonumber  \\ 
&&{\cal{F}} \left[ 
\frac{(x-iy)^2}{|\mathbf{r}|^5} \right] = -\frac{4\pi}{3}\,  e^{-i 2\beta} 
\sin^2\alpha    \,\,,
\label{sprytne_transformaty}
\eea  
From the numerical point of view calculations of ${\cal{H}}_{d11}$ and ${\cal{H}}_{d10}$ are the only change in (\ref{spinoroweRownaniaRuchu}) before we proceed to SOM algorithm.

\section{Accuracy tests for spinor dipolar condensates} 
\label{numerical_results}

\begin{figure}[h]
\centering
\includegraphics[scale=0.45]{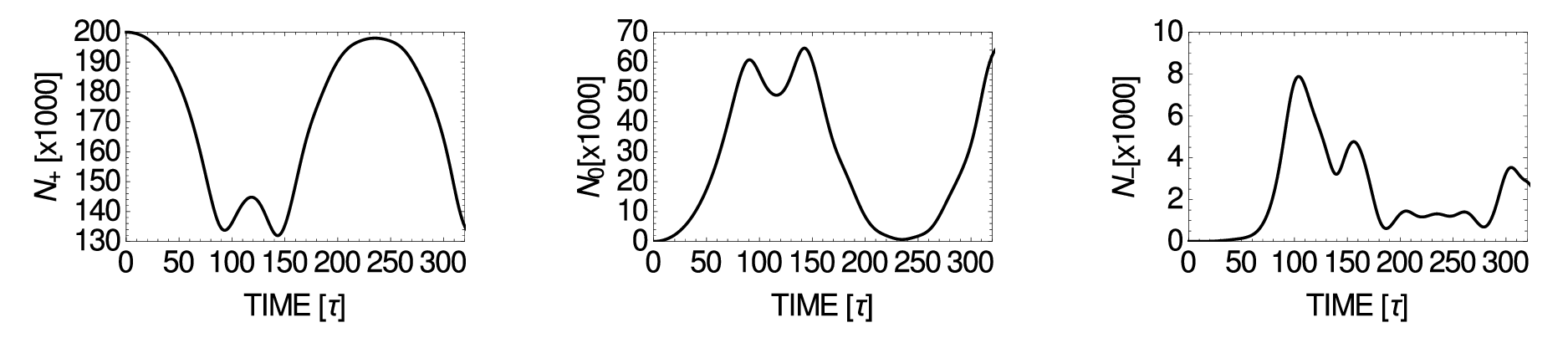}
\caption{Time evolution of the populations of all hyperfine states. Numerical parameters are as follows: $\Delta x=\Delta y=\Delta z=0.5 l$, $\Delta t=0.0005 \tau$, grid size $2^52^52^5$, $B_z=-40$ osc. units, $\gamma^2=0.0000257722$ osc. units. }
\label{fig_edh_norms}
\end{figure}

\begin{figure}[h]
\centering
\includegraphics[scale=0.5]{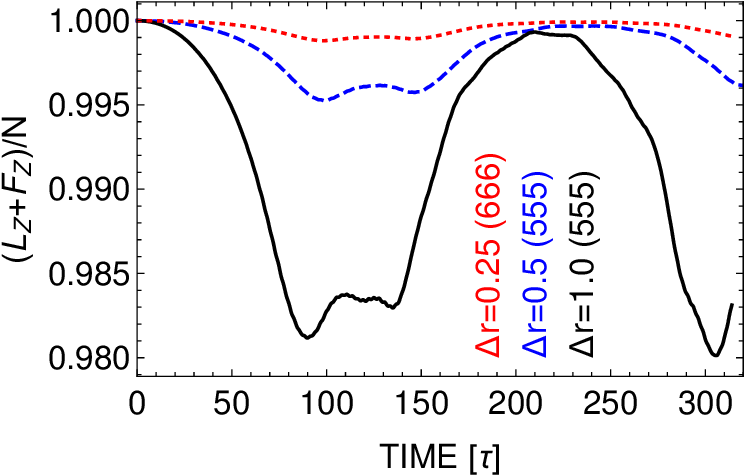}
\caption{Illustration of the conservation of the sum of the projections of the orbital angular momentum and the spin for various grid sizes. Figure shows the projection of the total angular momentum per atom in the system. Since we start calculations with all atoms being in the $m_F=+1$ Zeeman component, this quantity should be equal $1$ all the time. Here we used $\Delta t=0.0005 \tau$ and $\Delta x = \Delta y = \Delta z = \Delta r$. }
\label{fig_edh}
\end{figure}

\begin{figure}[h!]
\center
\includegraphics[width=14cm]{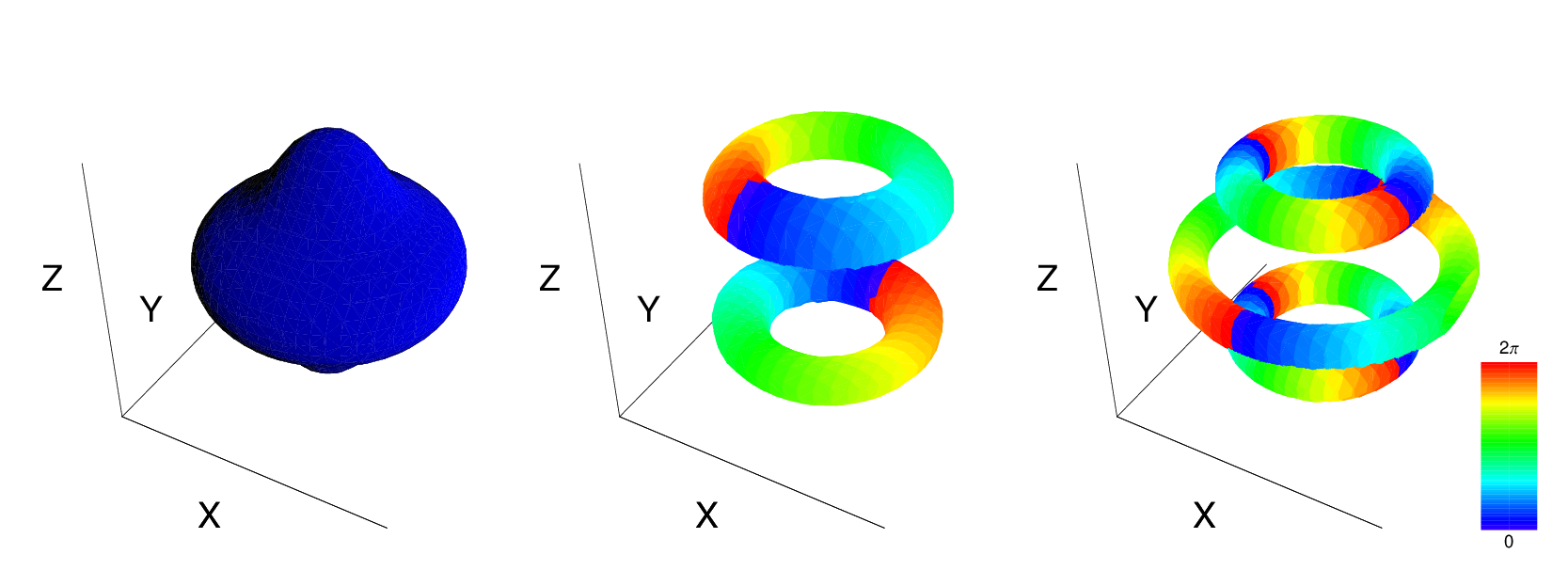}
\caption{Isodensity and phase of the order parameter at $t=0.14\,$s for the simulation presented in Fig. \ref{fig_edh_norms}. Values of the density for $m_F=1,0,-1$ (from left to right) equal $7.96\times10^{13}$ cm$^{-3}$, $7.96\times10^{13}$ cm$^{-3}$, and $7.96\times10^{12}$ cm$^{-3}$. The color on the surface shows the phase of the order parameter for a given component (color scale on the right). Note characteristic vortex structures in the $m_F=0$ component (it is a single quantized vortex) and in the $m_F=-1$ component (it is a doubly quantized vortex).}
\label{ladny3D}
\end{figure}

In Fig. \ref{fig_edh} we demonstrate how good is the SOM method with respect to the conservation of the sum of the total spin and the total orbital angular momentum. We consider a system of $N_{tot}=200000$ $\,^{87}$Rb atoms confined in a spherically symmetric harmonic trap ($\omega=2\pi \times 100$ Hz). Initially all the atoms populate the $m_F=+1$ Zeeman component. Then the magnetic field is turned on along the $z$ axis. Provided the value of the magnetic field is resonant \cite{EdH1}, the atoms start, due to dipolar forces, to flow to other components (see Fig. \ref{fig_edh_norms}). As it is seen, significant number of atoms is transferred from the $m_F=+1$ to $m_F=0$ and $m_F=-1$ states. Certainly, the $z$ projection of the total spin of the sample is changed during the evolution. However, Fig. \ref{fig_edh} clearly shows that the projection of the total angular momentum is conserved very well. 
In Tab. \ref{dipolartab} we present cumulative errors for the studied constants of motion for different spatial grids and time steps. All constants of motion, the projection of the total angular momentum (fourth column), the total energy (second column), and the total number of atoms (third column) are preserved better when the grid becomes finer and the time step gets smaller. As in the original Einstein-de Haas effect the atoms in $m_F=0$ and $m_F=-1$ components start to rotate around the direction of the magnetic field showing vortices in $m_F=0$ and $m_F=-1$ components (see Fig. \ref{ladny3D}).

\begin{table}[ht]
\centering
\begin{tabular}{c|c|c|c}\hline
$\Delta$t [$\tau$] & $\frac{1}{T}\int\! dt |\frac{\av{E}-E_0}{E_0}|\; [\times 10^{-5}]$ & $\frac{1}{T}\int\! dt
|\frac{N-N_0}{N_0}|\; [\times 10^{-5}]$ & $\frac{1}{T}\int\! dt\; |(L_z+F_z)/N|$ [$\hbar \omega \tau$]\\
   & (A)\;\;\;\;\;\;\;(B) & (A)\;\;\;\;\;\;\;(B)  &
(A)\;\;\;\;\;\;\;\;(B)  \\
\hline
0.001\phantom{00}   & $4.62415\;\;8.78206$  & $3.37224\;\;6.35519$     &
$0.998216\;\;0.991222$ \\
0.0005\phantom{0}   & $1.01042\;\;2.85040$  & $0.63968\;\;2.05545$     &
$0.999381\;\;0.991756$ \\
0.00025             & $0.41287\;\;0.54642$  & $0.17532\;\;0.39306$     &
$0.999832\;\;0.995903$ \\
\hline
\end{tabular}
\caption{Cumulative errors: (A) -- for the grid with $2^62^62^6$ points and the spatial steps $\Delta r =0.25 l$ (see Fig. \ref{fig_edh}); (B) -- for the grid with $2^52^52^5$ points and $\Delta r =0.5 l$. In both cases $T=314\tau$.}
\label{dipolartab}
\end{table}

\section{Degenerate Fermi gases} 
\label{Fermi_gases}

Ultracold Fermi gases serve themselves as ideal quantum simulators because of highly controllable laboratory conditions they can be realized at. Two-component Fermi gases have been recently used to study the properties of strongly interacting fermionic systems \cite{fermions_exp1,fermions_exp2}. It is challenging to confirm experimentally the occurrence of a ferromagnetic instability driven by the short range repulsion, responsible for the Stoner instability resulting in transition from para- to ferromagnetic phase.

It has been discovered already many years ago that the oscillations of the electron cloud in a many-electron atom can be viewed as a motion of a fluid characterized by the density and the velocity fields \cite{Wheeler}. Such motion is described by the hydrodynamic equations \cite{Wheeler}. It was already proposed by many authors and in a variety of contexts to use these equations to study a single-component Fermi gas (see for example \cite{Adhikari1,Salasnich,Karpiuk02,Karpiuk05}). Such a treatment was also applied to the superfluid Fermi gas of an equal mixture of spin-up and spin-down fermions in the BCS-BEC crossover (see for example \cite{Adhikari2}). However, as in a recent experiment \cite{fermions_exp2} also a metastable ferromagnetic phase due to strong repulsion between fermions in excited scattering states can be investigated. This metastable phase (as opposed to the ground state of spin-up and spin-down mixture which is formed of fermion pairs) is achieved by preparing initially the system in a configuration of two magnetic domain \cite{fermions_exp2}. To probe such a metastable phase, we propose to use the hydrodynamic equations \cite{Martin}. Assuming the velocity fields are rotation-free, the appropriate equations read 
\begin{eqnarray}
&&\frac{\partial}{\partial t} n_\pm = -\nabla(n_\pm\,\vec v_\pm),   \nonumber \\
&&m\frac{\partial}{\partial t}\vec v_\pm = -\nabla\left(\frac{\delta T}{\delta n_\pm}+\frac{m}{2}\vec v_\pm^2+V_{trap}+g\,n_\mp\right) ,
\label{hydrodynamics}
\end{eqnarray}
where $(n_j({\bf r},t), {\bf v}_j({\bf r},t))$ denote the density and velocity fields of $j-$th ($j=\pm$) component. $T$ is the intrinsic kinetic energy of the gas and is calculated as in the Thomas-Fermi approximation \cite{TFappr1,TFappr2}. Including the gradient corrections \cite{WeiKir1,WeiKir2} one gets
\begin{eqnarray}
\frac{\delta T}{\delta n_\pm} = A\,n_\pm^{2/3}-\xi\frac{\hbar^2}{2m}\frac{\nabla^2\sqrt{n_\pm}}{\sqrt{n_\pm}}  ,
\label{deltaT3D}
\end{eqnarray}
where $\xi=1/9$ and $A=6^{5/3}\hbar^2 \pi^{4/3}/(12m)$. Eqs. (\ref{hydrodynamics}) can be recast, by using the inverse Madelung transformation \cite{Madelung}, to the pseudo-Schr\"odinger equation
\begin{eqnarray}
i\hbar\frac{\partial}{\partial t}\psi_\pm &=& \left[-\frac{\hbar^2}{2m}\nabla^2+\frac{\hbar^2}{2m}(1-\xi)\frac{\nabla^2|\psi_\pm|}{|\psi_\pm|}   \right. \nonumber \\
&+&\left. A\,|\psi_\pm|^{4/3}+V_{trap}+g |\psi_\mp|^2  \right]\psi_\pm
\label{pseudopsi}
\end{eqnarray}
Eqs. (\ref{pseudopsi}) take the form of Eq. (\ref{spinorGP}), although here the system is described by two-component object (pseudo-wavefunction $(\psi_+({\bf r}),\psi_{-}({\bf r}))^T$) and therefore at each time step at each spatial point one has to diagonalize $2\times2$ matrix. Hence, the numerical method discussed in Sec. \ref{SOMspinor} can be directly used to study the dynamics of a Fermi mixture in the frame of Thomas-Fermi approximation.

In Ref. \cite{Martin} we investigated the ground state densities of repulsive two-component Fermi gases. Numerically, we just evolve the system according to Eqs. (\ref{pseudopsi}) by using the imaginary time technique. By increasing the strength of repulsion we observe the transition from the identical density profiles for two species towards, first, isotropic and, finally, anisotropic separations of two components (see Fig. 5 in Ref. \cite{Martin}). This indicates indirectly the existence of a ferromagnetic instability in a system of two repulsive Fermi gases. 

\begin{figure}[h!]
\center
\includegraphics[width=12cm]{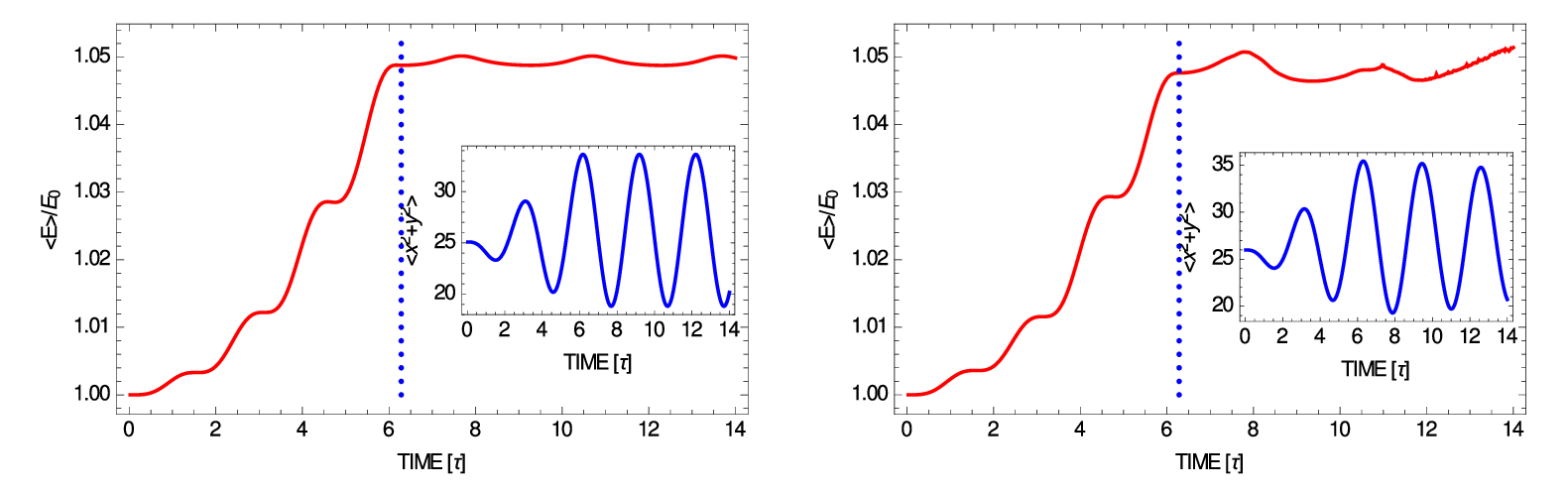}
\caption{Illustration of the conservation of the total energy for a system of two-component Fermi gas. The number of atoms in each component is $N_+=N_{-} =10$. The trap is initially disturbed (within the period up to the vertical dotted lines in each frame) by changing periodically the trap frequency. This pumps the energy into the system (main frames). When the trap is restored the energy is conserved at the level below one percent. The insets show the oscillations of each atomic cloud (both components behave in the same way) with the frequency which depends on the phase the system is in. The interaction strength changes from $g=7.0$ (paramagnetic phase, left frame) to $g=15.0$ (ferromagnetic phase, right frame). The grid size is $2^72^72^7$ ($2^62^62^6$) points for the left (right) frame and the time step is $\Delta t=0.0005 \tau$.  }  
\label{Fermi_gases_test}
\end{figure}

Here we investigate the accuracy of above discussed numerical algorithm after the trapping potential for two gases is periodically disturbed. In Fig. \ref{Fermi_gases_test} we show how preserved is the total energy of the system after the trap is restored for different values of the repulsion strength. For the set of $N_+=N_{-} =10$ atoms we analyze qualitatively different cases: the symmetric one corresponding to paramagnetic phase and the anisotropic separation case which represents the ferromagnetic phase (see Fig. 5 in Ref. \cite{Martin}). The frequency of the trap is modulated for a short period as $\omega (1+A\sin(\Omega t))$ with the small (just to avoid any nonlinear effects) driving amplitude $A=5\%$ and driving frequency $\Omega=2\, \omega$ ($\omega$ is the unperturbed trap frequency). So, we investigate the monopole oscillating mode. After the trapping potential is restored we observe oscillations of both components. The insets in Fig. \ref{Fermi_gases_test} show the oscillations of the following quantities: $\int (x^2+y^2) n_\pm(\bf r) d{\bf r}$ which are experimentally accessible after column density masurement along $z$ direction is done. For ferromagnetic phase the clouds oscillate with the frequency equals twice the trap frequency (the inset in the right frame in Fig. \ref{Fermi_gases_test}). This is because in this phase the atoms of different spins practically do not interact with each other -- the occupy opposite regions in the space (see Ref. \cite{Martin}). Therefore they behave as two independent ideal fermionic gases. On the other hand, for paramagnetic phase (left frame in Fig. \ref{Fermi_gases_test}) the frequency increases (and for $g=7.0$ osc. units equals $2.08\, \omega$) because during the oscillations atoms of different spins interact all the time.
In Tab. \ref{fermitab} we show cumulative errors for the studied constant of motion, i.e. for the total energy of the system after the trap is restored, for two spatial grids and the time step $\Delta t=0.0005 \tau$. Finer grid results in better conservation of the total energy.

\begin{table}[ht]
\centering
\begin{tabular}{c|c|c}\hline 
$\Delta$t [$\tau$] & $\frac{1}{T}\int\! dt |\frac{\av{E}-E_0^{\prime}}{E_0^{\prime}}|\; [\times 10^{-4}]$ & $\frac{1}{T}\int\! dt |\frac{\av{E}-E_0^{\prime}}{E_0^{\prime}}|\; [\times 10^{-4}]$ \\
   & (A)\;\;\;\;\;\;\;(B) & (A)\;\;\;\;\;\;\;(B) \\
\hline
0.0005          & $5.30668\;\;20.4478$  & $6.19379\;\;11.1891$   \\
\hline
\end{tabular}
\caption{Cumulative errors: (A) -- for the grid with $2^72^72^7$ points (spatial step $\Delta r=0.0937\, l$); (B) -- for the grid with $2^62^62^6$ points (and twice larger spatial step). Only period after the trap is returned to its initial shape is considered. Here, $E_0^{\prime}$ is the energy at time $t/\tau=2\pi$. The second (third) column presents results for the paramagnetic (ferromagnetic) phase.  }
\label{fermitab}
\end{table}

Even more sophisticated description of a Fermi system which involves the single-particle spin-orbitals can be rewritten making possible to use the procedure detailed in Sec. \ref{SOMspinor}. Let's see how it works. We assume that the many-body wave function of $N/2+N/2$ atoms is given by the single Slater determinant
\begin{eqnarray}
&&\Psi ({\bf x}_1,...,{\bf x}_{N})
= \frac{1}{\sqrt{N!}} \left |
\begin{array}{lllll}
\varphi_1({\bf x}_1) & . & . & . & \varphi_1({\bf x}_{N}) \\
\phantom{aa}. &  &  &  & \phantom{aa}. \\
\phantom{aa}. &  &  &  & \phantom{aa}. \\
\phantom{aa}. &  &  &  & \phantom{aa}. \\
\varphi_{N}({\bf x}_1) & . & . & . & \varphi_{N}({\bf x}_{N})
\end{array}
\right |    \nonumber  \\
\label{Slater}
\end{eqnarray}
with spin-orbitals: $\varphi_j({\bf x})=\varphi_j^{(1)}({\bf r})\, a(s)$ for $j=1,..,N/2$ and  \\ $\varphi_j({\bf x})=\varphi_{j-N/2}^{(2)}({\bf r})\, b(s)$ for $j=N/2+1,..,N$, which equally share the two spin states $a(s)$ and $b(s)$. The time-dependent Hartree-Fock equations for the spatial orbitals are then given by

\begin{eqnarray}
&& i\hbar \frac{\partial}{\partial t}  \varphi_j^{(1)} ({\bf r},t) =
 ( -\frac{\hbar^2}{2 m} \nabla^2 + V_{tr}({\bf r}) ) \; \varphi_j^{(1)} ({\bf r},t)    \nonumber  \\
&& + \sum_{k=1}^{N/2} \int d{\bf r^{\prime}}\, |\varphi_k^{(1)} ({\bf r}^{\prime},t)|^2\, V_{aa}({\bf r}-{\bf r}^{\prime})\;\;  \varphi_j^{(1)} ({\bf r},t)   \nonumber  \\
&& + \sum_{k=1}^{N/2} \int d{\bf r^{\prime}}\, |\varphi_k^{(2)} ({\bf r}^{\prime},t)|^2\, V_{ab}({\bf r}-{\bf r}^{\prime})\;\;  \varphi_j^{(1)} ({\bf r},t)   \nonumber  \\   
&& - \sum_{k=1}^{N/2} \int d{\bf r^{\prime}}\, (\varphi_k^{(1)} ({\bf r}^{\prime},t))^*\, V_{aa}({\bf r}-{\bf r}^{\prime})\, \varphi_j^{(1)} ({\bf r^{\prime}},t)\;\;  \varphi_k^{(1)} ({\bf r},t)
\label{HFeq}
\end{eqnarray}
for $j=1,...,N/2$, where the terms $V_{aa}$ and $V_{ab}$ describe the interactions between atoms being both in the $a(s)$ state and when one atom is in the $a(s)$ state and the other in the $b(s)$ one. Analogous set of equations is fulfilled by spatial orbitals $\varphi_j^{(2)} ({\bf r},t)$, $j=1,...,N/2$. The first and the second integrals and the last one are called Coulomb and exchange terms, respectively. Equations (\ref{HFeq}) for each kind of spatial orbitals can be written in the matrix form
\begin{eqnarray}
i \hbar \frac{\partial}{\partial t}\, \varphi^{(1)} =
(H_0 + V_{C}^{aa} + V_{C}^{ab} - V_{ex}^{aa})\, \varphi^{(1)}  \nonumber \\
i \hbar \frac{\partial}{\partial t}\, \varphi^{(2)} =
(H_0 + V_{C}^{bb} + V_{C}^{ba} - V_{ex}^{bb})\, \varphi^{(2)}    \,,
\label{HFmatrix} 
\end{eqnarray}
where $(\varphi^{(1)})^T=(\varphi_1^{(1)}({\bf r}),...,\varphi_{N/2}^{(1)}({\bf r}))^T$ and $(\varphi^{(2)})^T=(\varphi_1^{(2)}({\bf r}),...,\varphi_{N/2}^{(2)}({\bf r}))^T$. The Coulomb matrices are diagonal with equal elements, for example \\
$(V_C^{aa})_{jk}=\sum_{l} \int d{\bf r^{\prime}}\, |\varphi_l^{(1)} ({\bf r}^{\prime},t)|^2\, V_{aa}({\bf r}-{\bf r}^{\prime})\, \delta_{jk}$. The exchange matrices possess off-diagonal elements, for example $(V_{ex}^{aa})_{jk}=\int d{\bf r^{\prime}}\, (\varphi_k^{(1)} ({\bf r}^{\prime},t))^*\, V_{aa}({\bf r}-{\bf r}^{\prime})\, \varphi_j^{(1)} ({\bf r^{\prime}},t)$. Surprisingly, the dynamics of a many-fermion system is again described by the equation like Eq. (\ref{spinorGP}) in Sec. \ref{SOMspinor}. Hence, the method introduced in Sec. \ref{SOMspinor} can be used. Here, however, at each spatial point we have to diagonalize the square matrices of the size equal to the half of the number of atoms. Eqs. (\ref{HFmatrix}) in the very simple case when only contact interactions between different spin states is allowed (the effective Hamiltonians on the right-hand side of (\ref{HFmatrix})) are then diagonal) was already studied by us in Ref. \cite{Karpiuk04}. More demanding case when the pairing and ferromagnetic instabilities compete each other \cite{fermions_exp1,fermions_exp2} (and the Hamiltonian matrices in (\ref{HFmatrix}) are full) is under investigation \cite{Karpiuk17}.

\section{Conclusions}
\label{conclusions}

To summarize, we have presented the extended version of the SOM algorithm which turns out to be very efficient in simulating the evolution of spinor BEC systems with nonlocal interactions. We use this extension to solve the set of nonlinear partial integro-differential equations. In fact, this algorithm can be used to describe the evolution of any multicomponent system. It might be a spinor condensate of rubidium, chromium, erbium, or dysprosium atoms as well as the mixture of bosonic species. The algorithm can be also applied to multicomponent systems consisting of indistinguishable or distinguishable fermionic atoms. The combination of parallelization (with \textit{OpenMP} technique), the fast Fourier transform (via \textit{FFTW} routine), and the linear algebra algorithm for diagonalization (e.g., from \textit{lapack} packages) to the multicomponent SOM algorithm can be run very efficiently even on a typical desktop computer. We have proven that the algorithm conserves all constants of motion to a high accuracy.




\section*{Acknowledgments}
The work was supported by the (Polish) National Science Center Grant No. DEC-2012/04/A/ST2/00090.
Part of the results were obtained using computers at the Computer Center of University of Bialystok.

\end{document}